# Is Less *Really* More? Why Reducing Code Reuse Gadget Counts via Software Debloating Doesn't Necessarily Indicate Improved Security

Michael D. Brown, *Georgia Institute of Technology*     Santosh Pande, *Georgia Institute of Technology*

**Abstract**
Nearly all modern software suffers from bloat that negatively impacts its performance and security. To combat this problem, several automated techniques have been proposed to debloat software. A key metric used in many of these works to demonstrate improved security is code reuse gadget count reduction. The use of this metric is based on the prevailing idea that reducing the number of gadgets available in a software package reduces its attack surface and makes mounting a gadget-based code reuse exploit such as return-oriented programming (ROP) more difficult for an attacker.

In this paper, we challenge this idea and show through a variety of realistic debloating scenarios the flaws inherent to the gadget count reduction metric. Specifically, we demonstrate that software debloating can achieve high gadget count reduction rates, yet fail to limit an attacker's ability to construct an exploit. Worse yet, in some scenarios high gadget count reduction rates conceal instances in which software debloating makes security *worse* by introducing new, useful gadgets.

To address these issues, we propose a set of four new metrics for measuring security improvements realized through software debloating that are quality-oriented rather than quantity-oriented. We show that these metrics can identify when debloating negatively impacts security and be efficiently calculated using our static binary analysis tool, the Gadget Set Analyzer. Finally, we demonstrate the utility of these metrics in two realistic case studies: iterative debloating and debloater evaluation.

## 1. Introduction

Software debloating [1-5, 20, 24-27] is an emerging field of research focused on improving software security and performance by eliminating bloat that occurs as a byproduct of modern software engineering practices. While these practices enable the rapid development of complex, widely deployable, and feature-rich software, they produce software packages (programs, libraries, etc.) with large portions of code that are unnecessary in most end-use contexts. These portions of the package constitute software bloat and result in a variety of negative performance and security impacts [1, 6, 22, 23].

Software bloat affects virtually all software and primarily occurs vertically in the software stack across layers of abstraction [1]. Programs that depend on common shared code libraries such as libc typically only require a small number of functions provided by the library, but load the entire library into the program's memory space at runtime.

Software bloat also occurs laterally within software packages suffering from feature creep [23]. Examples include software such as cUrl, which can be used to transfer data via 23 different protocols, and iTunes, which features a media player, ecommerce platform, and hardware device interface within a single package. Since end users are unlikely to use every feature within these packages, the code associated with unused features contributes to software bloat.

Recently, several software debloating techniques [2-5, 20, 24-27] have been proposed that promise to improve software security by removing code bloat at various stages of the software lifecycle. One frequently utilized metric for measuring security improvements realized via debloating is the reduction in total count of code reuse gadgets available to an attacker, which we refer to as *gadget count reduction*. Several recent debloating publications [3-5, 24-27] claim their methods improve security citing gadget count reduction data as one form of evidence.

The relationship between gadget count reduction and improved security is based on the premise that reducing the total number of code reuse gadgets available in a software package reduces its attack surface. In turn, this decreases the likelihood of an attacker successfully constructing a code reuse exploit using techniques such as return, jump, or call-oriented programming (also known as ROP, JOP, and COP [7-9, 21]). At face value, gadget count reduction is an appealing security improvement metric as it is easily generated using existing automated static analysis tools [14], is quantifiable, and is directly relevant to a class of cyberattacks that have been the focus of intense research over the last decade [7-15].

The premise linking gadget count reduction to improved security holds only if the gadgets removed by debloating are critical to the construction of an exploit, and other gadgets with equivalent functionality are not available. For an attacker attempting to construct a code reuse exploit, the total number of gadgets available is irrelevant; what truly matters is whether or not the gadgets necessary to express their desired exploit and maintain malicious control flow are available. Recent research on gadget chaining tools [10, 11] and code reuse attack techniques [12, 13] have shown that attackers do not require a large, diverse, and fully expressive set of gadgets in order to craft an exploit. As a result, it is possible that debloating can achieve high gadget count reduction and indicate an improvement in security, yet fail to remove any of the gadgets an attacker needs to express and construct an exploit.

Even worse, our research indicates that debloating techniques that remove code from a package *introduce new gadgets* at a high rate as a side effect. Gadget count reduction data does not capture instances where a gadget removed via debloating is replaced by a newly introduced gadget. Except in rare cases where the total count of gadgets is increased by debloating, this poorly understood side effect is masked by the metric. This opens the possibility that debloating may "successfully" reduce the overall count of gadgets, all while introducing new, useful gadgets that have a negative impact on security.

### 1.1. Motivation

Despite these flaws, gadget count reduction is widely used as a security improvement metric in recent literature [3-5, 24-27] and has also been used a basis of comparison between two different debloating techniques [25]. We argue that the flaws inherent to this metric call into question claims of improved security or superiority of one debloating technique over another in these works. Our motivation in this work is to bring attention to the issues surrounding the use of gadget count reduction as a security improvement metric and suggest an alternative set of measures to counter this practice.

### 1.2. Paper Organization

In section 3 of this paper, we first present the results of our study of gadget introduction as a side effect of code-removing debloaters. We analyze the root causes of gadget introduction, and show that it is pervasive and occurs at a high rate using two different code-removing debloaters.

In section 4 of this paper, we propose a set of four new metrics for measuring the security impact of software debloating: *functional gadget set expressivity*, *functional gadget set quality*, *special purpose gadget availability*, and *gadget locality*. These metrics overcome the shortcomings of gadget count reduction by assessing the utility of the gadgets available to the attacker rather than the quantity.

In section 5, we demonstrate the practicality of these metrics by introducing our static binary analysis tool, Gadget Set Analyzer (GSA). GSA is capable of quickly and efficiently analyzing several program variants and calculating our proposed metrics in support of real-world debloating objectives.

In section 6, we use GSA to highlight the shortcomings of gadget count reduction and show the value of our proposed metrics in realistic debloating scenarios. In each scenario, positive gadget count reduction is achieved; however, GSA and our proposed metrics reveal that a significant number of scenarios are negatively impacted by gadget introduction.

Finally, in section 7 we demonstrate through two case studies that our metrics can be used to achieve meaningful security-oriented goals. In the first case study, we use GSA to identify and mitigate the negative security impacts of debloating in one of our scenarios through iterative debloating. In the second case study, we use GSA to conduct a side by side comparison of two debloating techniques and show that our metrics provide a more meaningful basis of comparison than gadget count reduction for determining which approach was more effective at improving security.

## 2. Background

In this section, we introduce necessary terms and review the current state of the art of relevant debloating techniques.

### 2.1. Relevant Terms

**Code Reuse Attacks:** Code reuse attacks are a class of attacks in which an attacker compromises the control flow of a program and redirects execution to an existing executable part of the program to cause a malicious effect, bypassing code injection defenses such as Write XOR Execute. In gadget-based code reuse attack methods such as ROP, JOP, and COP [7-9, 21], the attacker chains together short instruction sequences called gadgets present in the program in a specific order to construct a malicious payload without injecting code.

**Gadget**: A gadget suitable for use in a code reuse attack is a short sequence of machine instructions that end in a return, indirect jump, or indirect call instruction. Gadgets can be chained together using the control flow properties of the terminating instruction to create a malicious payload comprised entirely of *existing* code segments.

**Gadget Types:** When constructing a gadget chain, gadgets are used for one of two purposes. Functional gadgets are used as abstract instructions to express the attacker's malicious intent. Gadgets that can be used to perform important non-expressive actions such as invoking system calls or maintaining gadget chain control flow are called special purpose gadgets.

**ROP/JOP/COP Methods**: In ROP [7], the attacker takes control of the stack and chains gadgets ending in return instructions together to create a malicious payload. The attacker maintains malicious control flow by exploiting the behavior of the stack to direct execution to the next gadget in the chain (triggered when the return instruction terminating the current gadget is executed).

In JOP [8, 21], the attacker chains gadgets ending in unconditional jump or call instructions together to create a malicious payload. This approach is not dependent on the stack and return instructions for control flow, and as such circumvents many ROP defenses. To initialize the attack and control the flow of execution from gadget to gadget, JOP relies on special purpose gadgets. A critical special purpose gadget, the dispatcher, maintains an ordered table of functional gadgets and ensures each functional gadget's jump or call targets the dispatcher gadget after execution.

COP [9] is a specialized variant of JOP, in which functional gadgets are limited to sequences ending in call instructions only. Gadget to gadget control flow in COP is accomplished using a dispatcher gadget similar to a JOP dispatcher, or through the use of other special purpose gadgets such as trampoline gadgets.

### 2.2. Related Work: Code-removing Debloaters

**CHISEL:** Lee et al. [3] recently proposed an automatic method for debloating unnecessary features from program source code called CHISEL. CHISEL takes as input a specification script that outputs whether or not a debloated variant satisfies the desired program properties. Using an iterative, feedback-directed program reduction algorithm, CHISEL progressively removes segments of the program that are not necessary to satisfy the desired properties.

This work cites gadget count reduction data as evidence of security improvement through attack surface reduction, but does not provide further analysis of the gadgets present in their debloated programs. CHISEL's source code and benchmarks have been made publicly available [16, 17].

**TRIMMER:** Sharif et al. [4] recently proposed an automated method for debloating unnecessary functionality from software named TRIMMER. TRIMMER takes as input a static user defined configuration that expresses the deployment context for a particular program. Static configuration data is treated as a compile time constant and is propagated throughout the program. This is followed by custom, aggressive compiler optimizations to prune functionality from the program.

The authors provide gadget count reduction data as evidence that TRIMMER reduces the attack surface of a program by removing exploitable gadgets; However, they provide no explanation of what makes a gadget exploitable as opposed to non-exploitable. Additionally, their data indicates that syscall gadgets were introduced as a result of debloating, yet no explanation or investigation of this occurrence is provided. TRIMMER has not yet been made publicly available.

**CARVE:** Brown and Pande [20] recently proposed an automated feature debloating technique, named CARVE. CARVE requires users (software developers) enrich their source code with feature mappings that identify code related to one or more debloatable features. These mappings can optionally specify replacement code that prevents violation of high-level security policies. CARVE takes the annotated source code and a debloating specification as input to a specialized source code pre-processor pass that scans the enriched source code for corresponding feature mappings. When found, CARVE performs syntax-aware analysis on the mapped code and intelligently removes it in a sound manner. The debloated source code is then compiled to produce a debloated binary using the same build process as the original version. CARVE has been made available by the original authors for use in this work.

**RAZOR:** Qian and Hu et al. [25] recently proposed an automated feature debloating technique for binaries called RAZOR. RAZOR takes as input a program binary and set of test cases that exercise desired functions, and performs a three-stage process that identifies desired code via dynamic tracing, uses heuristics-based analysis to identify additional function related code that was not exercised by the test cases, and synthesizes a new debloated binary from the identified code.

The authors use gadget count reduction data as one of two security improvement metrics (the other is known vulnerability elimination) to show security improvements with their approach. Additionally, gadget count reduction data is used to directly compare the efficacy of RAZOR against that of CHISEL. This work does not provide further analysis of the gadgets present in their debloated programs or CHISEL's programs. As of the writing of this paper, RAZOR has not yet been made publicly available.

## 3. Gadget Introduction via Debloating

Techniques such as CHISEL, TRIMMER, CARVE, and RAZOR that debloat by altering a software package's representation (source code, intermediate representation, or binary) can introduce new gadgets into the debloated variant in a manner that is difficult to predict. Since we do not have prior knowledge of which gadgets are useful to an attacker, gadget introduction can potentially offset security improvements realized through debloating, or even make a debloated package *less secure*. We describe the root causes of gadget introduction in the following two sections.

### 3.1. Introduction of Intended Gadgets

During compilation, source code is translated to an intermediate representation and optimized before the resulting binary is created during code generation. Gadgets comprised of compiler generated binary instructions are referred to as *intended gadgets*. In source code and compiler-based debloating techniques (CHISEL, TRIMMER, and CARVE), changes to package's representation caused by debloating can cause downstream compiler stages to make different optimization and code generation decisions. Though RAZOR debloats packages in their binary format, it still causes significant changes to the package representation as a result of its binary re-synthetization approach. Regardless of the mechanism by which the binary instructions composing a software package are changed, the changes cause the introduction of new intended gadgets.

Consider the control flow graph (CFG) excerpts from `libcurl` shown in Figure 1. The excerpt on the left is from the original version of the function `curl_version`, and the excerpt on the right shows the results of debloating eight lines from the corresponding source code. Binary instructions present in both versions are in **bold** text, and the return instruction that forms ROP gadgets is in ***bold italicized*** text. Removing the source code results in fewer binary instructions as expected, however this shorter sequence of instructions has simpler control flow. Specifically, the `jmp` instruction on line 19 and all but one instruction in the basic block following it are removed. As a result, all three basic blocks in the original version are merged into a single basic block in the debloated version. This change in control-flow increases the range of instructions the compiler can reorder to maximize performance (lines 24-27 in the original version versus lines 1-11 in the debloated version). Comparing the number of unique intended ROP gadgets produced by these two sequences, the net result is an increase in the gadget count. Two gadgets, [pop rbx; pop rbp; retn;] and [pop rbp; retn;], are present in both versions. One gadget is eliminated from the original version (lines 24-27), but two gadgets are introduced in the debloated version (lines 7-11 and 8-11).

**Figure 1:** Control flow graph snippets of `curl_version`.

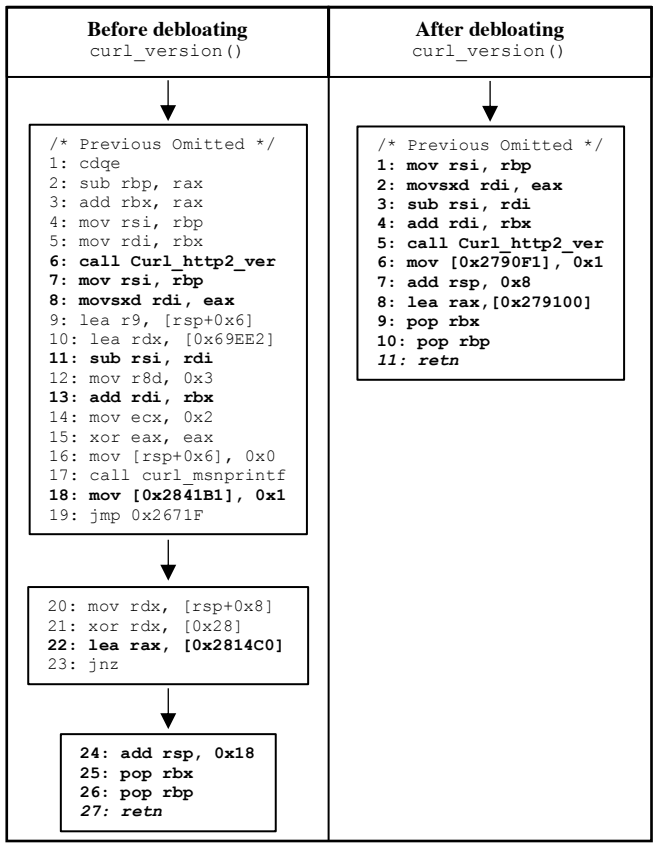

In addition to instruction reordering, we have observed a number of other compiler operations that can be triggered by debloating (this is not an exhaustive list):

- Eliminating branches (especially those close to an indirect branch or a procedure boundary) can alter control flow and introduce new gadgets.
- Debloating entries in static arrays can trigger or prevent compile time optimizations for loops that iterate through these arrays. This has a significant effect on the code the compiler generates and the control-flow mechanisms used to implement the loop.
- Inline function expansion can be triggered by debloating function call sites, altering control flow and potentially introducing new gadgets.
- Debloating can trigger dead code elimination across basic blocks, resulting in changes to control flow such as the merging of basic blocks that can introduce new gadgets.
- Debloating can result in certain program variables becoming statically determinable, triggering constant propagation optimizations.
- Debloating can increase opportunities for code motion-based optimizations such partial redundancy elimination and loop invariant code motion.

### 3.2. Introduction of Unintended Gadgets

For ISAs with variable length instructions such as x86 and x86-64, it is possible to decode instructions from an offset other than the original instruction boundary to obtain new, potentially valid instruction sequences [7]. Code reuse gadgets found in these sequences are referred to as *unintended gadgets*. When debloating alters compiler generated instructions, it also introduces new unintended gadgets when the altered code is interpreted at unintended offsets. Even in simple cases where a small number of intended instructions are removed, a significant number of new unintended gadgets are potentially introduced due to reinterpretation of code from unintended offsets.

For example, consider the sequence of instructions taken from the last basic block of `curl_version` before debloating, shown in the top segment of Figure 2. Unintended gadgets can be identified by interpreting the sequence at an offset of one byte and three bytes after the first instruction boundary, shown in the bottom two segments of Figure 2. After debloating, the instructions in the final basic block (top segment of Figure 3) have changed significantly. In addition to altering the intended gadgets found in this basic block, debloating also changes the unintended gadgets found at various offsets from the original instruction boundary (bottom two segments of Figure 3). These unintended gadgets differ significantly from those found in the original sequence. The net result of this operation is the elimination of the unintended gadgets in the original sequence and the introduction of the unintended gadgets in the resulting sequence.

**Figure 2:** Final basic block of `curl_version` before debloating, interpreted at three different byte offsets.

```
48 83 c4 18:    add rsp,0x18
5b:             pop rbx
5d:             pop rbp
c3:             ret

83 c4 18:       add esp,0x18
5b:             pop rbx
5d:             pop rbp
c3:             ret

18 5b 5d:       sbb BYTE PTR[rbx+0x5d],bl
c3:             ret
```

### 3.3. Prevalence of Gadget Introduction

To determine the degree to which software debloating introduces new gadgets, we debloated a variety of common software packages at varying levels of aggressiveness with two different code-removing debloaters. We analyzed each package and its variants using ROPgadget [14] to catalog its gadgets by type, and then performed a set-wise comparison to identify gadgets present in the debloated variants that were not present in the original package.

For this analysis we used CARVE and CHISEL, which were the only code-removing debloating tools available. We debloated four software packages with CARVE from the author provided benchmark set [20]: `libmodbus` v3.1.4, `Bftpd` v4.9, `libcurl` v7.61, and `mongoose` v.6.8. Each package was debloated at three levels of aggressiveness, defined below:

- Conservative (C): Some peripheral features in the package are removed.
- Moderate (M): Some peripheral features and some core features are removed from the package.
- Aggressive (A): All debloatable features except for a small set of core features are removed from the package.

These feature sets were selected to reflect reasonable real-world use cases. Detailed feature set information for each scenario is included in Appendix A.

We used CHISEL to debloat nine software packages from the author provided benchmark set [17]: `bzip2` v1.05, `chown` v8.2, `date` v8.21, `grep` v2.19, `gzip` v1.2.4, `mkdir` v5.2.1, `rm` v8.4, `tar` v1.14, and `uniq` v8.16. We used the author provided specifications (roughly equivalent to the above definition of aggressive debloating) to create debloated variants. Detailed information on these debloating specifications can found on the author's benchmark repository [17].

All of the software packages and their debloated variants were built on the same platform, a virtual machine running the 64-bit Ubuntu 18.04.1 LTS. For consistency, we used the same compilers used by the original authors to build their

**Figure 3:** Final basic block of `curl_version` after debloating, interpreted at three different byte offsets.

```
c6 05 0f 4d
24 00 01:       mov BYTE PTR[rip+0x244d0f],0x1
48 83 c4 08:    add rsp,0x8
48 8d 05 13
4d 24 00:       lea rax,[rip+0x244d13]
5b:             pop rbx
5d:             pop rbp
c3:             ret

08 48 8d :      or BYTE PTR [rax-0x73],cl
05 13 4d 24 00: add eax,0x244d13
5b:             pop rbx
5d:             pop rbp
c3:             ret

13 4d 24:       adc ecx,DWORD PTR [rbp+0x24]
00 5b 5d:       add BYTE PTR[rbx+0x5d],bl
c3:             ret
```

respective benchmarks: GCC version 7.3.0 for CARVE packages, and clang version 7.0.1 for CHISEL packages.

Table 1 contains the results of our gadget count reduction and gadget introduction analysis. As shown in first column, debloating successfully reduced the total count of gadgets in all scenarios, and in most cases the total count is reduced by a large degree (>15%).

The introduction of new gadgets via debloating is neither a rare nor limited occurrence. In all scenarios, a significant portion of the functional gadgets remaining after debloating were introduced gadgets (35% was the *smallest* observed rate). In 12 of 21 scenarios, introduced gadgets accounted for the majority of remaining gadgets. Further, gadget introduction occurs at a significant rate across all gadget subcategories (ROP, JOP, COP), aggressiveness levels, benchmarks, and debloating tools.

Significant levels of gadget introduction were also observed for special purpose gadgets (a value of N/A indicates that no gadgets of this type were present in the variant). In 14 of the 21 scenarios, debloating resulted in the introduction of new special purpose gadgets. As was the case with functional gadgets, special purpose gadget introduction is not strongly correlated to debloating aggressiveness, gadget subcategory, benchmark type, or debloating tool used.

The prevalence of gadget introduction has serious implications for the use of gadget count reduction as a security metric. Our data strongly indicates that the gadgets present in a debloated variant will not be a proper subset of the gadgets in the original package, rendering metrics that capture only the change in size of a set insufficient and superficial.

## 4. Gadget Utility Metrics

In this section, we propose a set of four new metrics for assessing the security impact of software debloating. Our metrics are designed to measure the degree to which debloating adversely affects the construction of gadget-based code reuse exploits by assessing the *utility* of the gadgets present after debloating as opposed to the *quantity*. Since functional gadgets and special purpose gadgets are utilized in different manners, we propose metrics suited for each gadget type. Additionally, we propose a metric that addresses the residual utility of gadgets that are unaffected by debloating.

Table 1: Number and percentage of gadgets introduced by type in debloated software packages.

| | Debloated Variant | Gadget Count Reduction | Introduced Functional Gadgets | | | | Introduced Special Purpose (S.P.) Gadgets | | |
|---|---|---|---|---|---|---|---|---|---|
| | | | All Gadgets | ROP Gadgets | JOP Gadgets | COP Gadgets | System Call Gadgets | JOP Specific S.P. Gadgets | COP Specific S.P. Gadgets |
| CARVE | libmodbus (C) | 89 (14%) | 222 (39%) | 149 (33%) | 73 (60%) | 33 (66%) | N/A | 2 (67%) | N/A |
| | libmodbus (M) | 108 (16%) | 221 (40%) | 163 (37%) | 58 (55%) | 35 (67%) | N/A | 0 (0%) | N/A |
| | libmodbus (A) | 143 (22%) | 252 (49%) | 156 (41%) | 96 (73%) | 44 (73%) | N/A | 0 (0%) | N/A |
| | Bftpd (C) | 40 ( 5%) | 249 (35%) | 202 (32%) | 47 (61%) | 20 (80%) | N/A | 0 (0%) | N/A |
| | Bftpd (M) | 124 (16%) | 260 (41%) | 197 (36%) | 63 (74%) | 44 (90%) | N/A | 0 (0%) | N/A |
| | Bftpd (A) | 220 (29%) | 196 (37%) | 167 (34%) | 29 (62%) | 14 (78%) | N/A | 0 (0%) | N/A |
| | libcurl (C) | 214 (2%) | 4727 (51%) | 1858 (45%) | 2864 (56%) | 2165 (52%) | 5 (100%) | 17 (30%) | 305 (99%) |
| | libcurl (M) | 1470 (15%) | 4178 (52%) | 1631 (44%) | 2547 (58%) | 2003 (56%) | 0 (0%) | 10 (23%) | 287 (99%) |
| | libcurl (A) | 3766 (40%) | 3334 (58%) | 1422 (49%) | 1911 (68%) | 1664 (69%) | 4 (100%) | 5 (26%) | 171 (99%) |
| | Mongoose (C) | 18 (2%) | 412 (33%) | 307 (29%) | 105 (66%) | 55 (78%) | N/A | 0 (0%) | N/A |
| | Mongoose (M) | 52 (4%) | 396 (33%) | 277 (27%) | 119 (60%) | 63 (80%) | N/A | 0 (0%) | 1 (100%) |
| | Mongoose (A) | 99 (8%) | 405 (35%) | 258 (27%) | 147 (67%) | 63 (80%) | N/A | 0 (0%) | N/A |
| CHISEL | bzip2 | 442 (65%) | 152 (64%) | 99 (59%) | 53 (76%) | 45 (75%) | N/A | 1 (100%) | N/A |
| | chown | 327 (65%) | 94 (54%) | 68 (47%) | 26 (84%) | 14 (87.5%) | N/A | 1 (100%) | N/A |
| | date | 260 (55%) | 119 (56%) | 89 (51%) | 30 (77%) | 23 (85%) | N/A | 1 (100%) | N/A |
| | grep | 378 (36%) | 479 (72%) | 380 (69%) | 99 (87%) | 79 (93%) | N/A | 1 (100%) | N/A |
| | gzip | 195 (46%) | 156 (68%) | 127 (65%) | 29 (83%) | 25 (86%) | N/A | 1 (100%) | N/A |
| | mkdir | 101 (48%) | 47 (42%) | 35 (38%) | 12 (67%) | 5 (83%) | N/A | 1 (100%) | N/A |
| | rm | 384 (72%) | 77 (52%) | 47 (41%) | 30 (91%) | 15 (88%) | N/A | 1 (100%) | N/A |
| | tar | 1355 (84%) | 144 (57%) | 75 (44%) | 69 (86%) | 52 (88%) | N/A | 2 (100%) | N/A |
| | uniq | 176 (59%) | 53 (43%) | 33 (34%) | 20 (77%) | 5 (71%) | N/A | 1 (100%) | N/A |

### 4.1. Functional Gadget Set Expressivity

Functional gadgets are used as abstract instructions that perform basic computational operations such as addition, register loading, and logical branching to construct a malicious payload. The *expressivity* of a set of gadgets is a measure of the computational power the set of gadgets permit.

The expressivity of a set of gadgets is typically measured against the bar of Turing-completeness. A set of gadgets is considered Turing-complete if it is sufficient to express any arbitrary program, i.e. it is computationally universal. While this level of expressivity is not difficult to achieve in practice [7-9, 12], it does not represent the minimum level necessary to construct a practical ROP exploit. Exploits that mark a region of memory as writable, inject malicious code, and redirect execution to this injected code do not require Turing-complete levels of expressivity [13].

For a gadget set to achieve a certain level of expressivity, it must contain at least one gadget supporting each necessary computational class. Thus, a straightforward measure of the expressivity of a gadget set is the number of computational classes satisfied by the gadget set. For example, a gadget set that can be used for addition, register loading, and conditional branching is considered more expressive than one that supports only addition and register loading.

If debloating removes all gadgets that perform a certain computation, the number of satisfied classes decreases. As a result, an attacker may not be able to express their desired exploit. If debloating introduces gadgets that perform previously unavailable computations, the number of satisfied classes increases, indicating a negative security impact.

### 4.2. Functional Gadget Set Quality

When selecting a functional gadget to perform a particular computational task, the attacker prefers to use gadgets that perform the required task and nothing more. However, given the nature of gadget-based code reuse attack methods and the unpredictable nature by which gadgets can be found in software packages, this is not always possible. In order to perform a specific computational task such as moving a value into a specific general purpose register, the attacker may be forced to use a functional gadget that contains extraneous instructions. Such instructions may interfere with the overall exploit chain in a variety of ways (not an exhaustive list):

- If an extraneous instruction writes to a memory location that is calculated at runtime, the gadget chain must ensure that the calculated memory location points to writeable memory or a chain-breaking segmentation fault may occur.
- If an extraneous instruction performs a conditional branch, the gadget chain must ensure that the appropriate condition flag is set to ensure the desired branch is taken during exploit execution.

- If an extraneous instruction stores a value in a register, the gadget chain must ensure this operation does not overwrite a necessary value placed in this register by a previous gadget in the chain.
- If an extraneous instruction in a ROP gadget modifies the stack pointer register value this must be considered during payload construction to ensure that the stack pointer points to the next gadget after executing the current one.
- For JOP and COP gadgets, if an extraneous instruction overwrites the indirect jump/call target register, the gadget chain must ensure that the address of the next gadget or appropriate special purpose gadget is stored into the register, which may not be possible.

Follner et al. [18] first proposed a gadget quality metric for scoring ROP gadgets according to the relative impact of these side effects. In their work, Follner et al. use this metric to assess the impact of building software with support for MPX CPU extensions, however their scoring mechanism is sufficiently general to apply to any software transformation, including software debloating. For use in our work, we have extended the original scoring mechanism and tool implementation (called Gality) proposed by Follner et al. to account for side effects not originally considered in their work (the presence of conditional branches and non-static adjustments to the stack pointer value) and to score JOP and COP gadgets (their work focused on ROP gadgets only).

The extended scoring method we use to calculate functional gadget set quality first eliminates gadgets that have side effects that make them impossible to use in a gadget chain. Next, a score is assigned to each gadget indicating its quality. Scores are calculated by starting with a value of 0, increasing the score for each detected side effect. The higher the score, the more difficult the gadget is to use in an exploit chain. The gadget elimination criteria and a full list of side effects and their associated scores detected by our extended version of Gality are included in Appendix B.

To determine the impact of debloating on functional gadget set quality, we examine the number of useful gadgets (i.e. gadgets not eliminated by the scoring method) and the average quality score of the useful gadgets. If debloating eliminates all of the useful gadgets or causes a significant increase in the average gadget quality score for a particular attack method (ROP/JOP/COP), then a positive security impact has been achieved. If debloating causes a significant decrease in the average gadget quality score, this indicates a negative security impact.

### 4.3. Special Purpose Gadget Availability

Special purpose gadgets are used to perform important non-expressive actions in an exploit gadget chain. These gadgets must meet specific criteria, and are typically encountered infrequently. Without special purpose gadgets, some exploits are not possible. For example, JOP/COP exploits do not use the stack, and instead rely on special purpose gadgets such as dispatchers and trampolines to maintain control flow from one gadget to the next. Also, exploits that must invoke system calls require special purpose gadgets.

Given their importance, the availability of special purpose gadgets is a useful metric for determining if debloating has reduced the types of exploits an attacker can construct. The availability of special purpose gadgets can be measured by maintaining counts of each type of special purpose gadget. If a debloating operation removes all of the special purpose gadgets of a particular type, then the attacker may not be able to construct their desired exploit. This metric is also capable of detecting the negative side effects of gadget introduction. If debloating introduces new special purpose gadgets of a particular type, this increase in availability is observable.

### 4.4. Gadget Locality

Code-reuse attacks are complex and require significant effort on the attacker's part to construct. The attacker must first discover a suitable attack vector (typically an exploitable memory corruption vulnerability) on the target system, obtain a copy of vulnerable binary, scan it for gadgets, and use those gadgets to construct their exploit. The final output of this resource intensive process is a malicious payload that consists of an ordered chain of offsets where the necessary gadgets are located in the vulnerable binary and any required data. Despite the high effort required to construct these exploits, they are fragile. If the gadgets in the exploit chain have been modified or have been moved to a new offset by a software update or patch, then the exploit is very likely to fail. As a result, code reuse exploits are typically not portable from one variant of a software package to the next. Attackers rely on software homogeneity to target multiple systems with the same code reuse exploit; if they encounter a variant of a target program they must start the exploitation process anew.

Software debloating prevents the attacker from utilizing gadgets present in bloat code, however it can also have a scrambling effect on the resulting debloated binary that relocates or alters gadgets that are not removed (or introduced). This effect increases software diversity, especially for feature-removing debloaters that are end user driven, as opposed to techniques that are focused on removing unreachable library code [5, 26, 27] which cannot be used to produce a large number of package variants. Increasing software diversity improves security as a form of moving target defense. Specifically, if an attacker develops a code reuse exploit for a software package, the exploit is not likely to be portable to a debloated version of that package if the debloating process removed, altered, or relocated a sufficiently large percentage of the gadgets present in the binary. To exploit the debloated variant, the attacker would need to acquire the debloated binary, discover its gadgets, and attempt to rewrite and re-package their exploit.

We define the *gadget locality* of a debloated package variant as the percentage of gadgets remaining after debloating that have not been altered and can be found at the same offset as the original package. If debloating a package results in 0% gadget locality, then a code reuse exploit using gadgets found in the original package cannot be readily used to exploit the debloated version. If debloating results in 100% gadget locality, then an exploit targeting the original package can be used against the debloated version provided the exploit chain does not contain a gadget removed during debloating.

### 4.5. Discussion of Metrics

We do not claim that these metrics are comprehensive. We encourage further discussion that will lead to the exploration of additional useful security-oriented metrics for debloating. Other measures of gadget utility with respect to specific attack methods (code injection via a ROP chain invoking `VirtualProtect` on Windows 7 x64 systems) have been explored by Follner et al. [18], which are potentially applicable to the problem of measuring the security impact of debloating in less generalized contexts.

## 5. Gadget Set Analyzer (GSA)

To assess the practicality of our metrics, we created a static binary analysis tool capable of analyzing a software package and its debloated variants to capture changes in functional gadget expressivity and special purpose gadget availability.

### 5.1. Operation

GSA operates in a straightforward manner and makes use of existing tools where possible. GSA takes as input the original package binary and one or more debloated variants. First, GSA uses ROPgadget [14] to search each binary for unique ROP, JOP and Syscall gadgets. GSA then performs secondary search of these results to identify unique COP gadgets and JOP/COP specific special purpose gadgets.

Once GSA has compiled completed its secondary search, it runs our extended version of Gality on the gadgets found in each package variant to calculate the number of useful gadgets, individual quality scores for these gadgets, and the average quality score across the set of useful gadgets.

Next, GSA utilizes a tool first proposed and implemented by Homescu et al. [12] for classifying short ROP gadgets (called microgadgets) into computational classes. This microgadget scanner then determines the expressive power of the gadget set relative to different levels of expressivity by analyzing the number of computational classes satisfied. GSA uses this scanner to analyze the gadget set with respect to three levels of expressivity originally proposed in [12]: simple Turing-completeness, expressivity required for practical ROP exploits, and expressivity required for practical, ASLR-proof ROP exploits. Each level requires a different number of computational classes be satisfied by at least one gadget: 17 for simple Turing-completeness, 11 for practical ROP exploits, and 35 for ASLR-proof, practical ROP exploits.

After detecting special purpose gadgets and collecting functional gadget set quality and expressivity data, GSA compares the data for each debloated variant against the data collected on the original binary to calculate changes in functional gadget set expressivity, functional gadget set quality, and special purpose gadget availability. Finally, GSA attempts to match the location and content of each gadget present in the debloated binary to one of the gadgets in the original binary. Successful matches are recorded and used to calculate gadget locality.

GSA also provides a means for the user to address gadgets introduced by debloating. Using the binary analysis library angr [19], GSA attempts to identify the names of functions containing introduced gadgets. This information can be used to identify the source of an undesirable debloating operation, potentially allowing the user alter their debloating specification to generate a new variant that does not introduce the gadget. GSA provides this as a "best effort" feature, and cannot guarantee that a function name can be retrieved. This is due to several factors including imprecision associated with generating a CFG in angr and the mechanism by which the gadget was introduced.

### 5.2. Limitations

GSA's calculation of functional gadget set expressivity has the same limitations as the gadget expressivity scanner it incorporates. Specifically, the microgadget scanner used by GSA generates functional gadget expressivity data based solely on an analysis of short gadgets in the set. As a result, a gadget set may be more expressive than reported by GSA if longer gadgets excluded from analysis can satisfy additional computational classes. Also, the microgadget scanner only calculates the expressivity of ROP gadget sets, and does not calculate expressivity for JOP or COP gadget sets.

Additionally, dynamically linked external libraries are not scanned by GSA. At runtime, these libraries are loaded into memory and their code is mapped to the package's address space. Gadgets present in these libraries contribute to overall expressivity and special purpose gadget availability, and also should be considered for a holistic view.

### 5.3. Performance

GSA and its dependencies perform binary analysis exclusively with static techniques. As such, the time required to run GSA increases with the size of the binary. GSA is performant, typically requiring less than 30 seconds to analyze a binary and three variants on a typical laptop. The maximum time required to execute GSA we observed was 2 minutes and

Table 2: Computational Classes Satisfied (and Reduction) by Gadget Set for Three Functional Expressivity Levels.

| CARVE | | | | CHISEL | | | |
|---|---|---|---|---|---|---|---|
| Package Variant | Practical ROP Exploit Classes | ASLR-Proof ROP Classes | Simple Turing Complete Classes | Package Variant | Practical ROP Exploit Classes | ASLR-Proof ROP Classes | Simple Turing Complete Classes |
| libmodbus (C) | 6 of 11 (0) | 10 of 35 (3) | 6 of 17 (1) | bzip2 | 3 of 11 (2) | 5 of 35 (2) | 1 of 17 (2) |
| *libmodbus (M)* | *6 of 11 (0)* | *13 of 35 (0)* | *7 of 17 (0)* | chown | 3 of 11 (2) | 5 of 35 (4) | 2 of 17 (2) |
| *libmodbus (A)* | *6 of 11 (0)* | *13 of 35 (0)* | *7 of 17 (0)* | date | 3 of 11 (2) | 5 of 35 (4) | 2 of 17 (2) |
| **Bftpd (C)** | **7 of 11 (-1)** | 12 of 35 (3) | 6 of 17 (1) | grep | 3 of 11 (2) | 6 of 35 (5) | 2 of 17 (5) |
| **Bftpd (M)** | **7 of 11 (-1)** | **17 of 35 (-2)** | 6 of 17 (1) | gzip | 3 of 11 (2) | 5 of 35 (1) | 1 of 17 (2) |
| Bftpd (A) | 6 of 11 (0) | 11 of 35 (4) | 5 of 17 (2) | mkdir | 3 of 11 (0) | 6 of 35 (0) | 1 of 17 (1) |
| **libcurl (C)** | 9 of 11 (0) | **26 of 35 (-1)** | **11 of 17 (-1)** | rm | 3 of 11 (2) | 5 of 35 (4) | 2 of 17 (2) |
| libcurl (M) | 9 of 11 (0) | 24 of 35 (1) | 10 of 17 (0) | tar | 3 of 11 (2) | 5 of 35 (4) | 1 of 17 (4) |
| **libcurl (A)** | **10 of 11 (-1)** | 24 of 35 (1) | 10 of 17 (0) | uniq | 3 of 11 (0) | 5 of 35 (4) | 1 of 17 (3) |
| Mongoose (C) | 7 of 11 (0) | 10 of 35 (6) | 8 of 17 (0) | | | | |
| Mongoose (M) | 7 of 11 (0) | 10 of 35 (6) | 8 of 17 (0) | | | | |
| Mongoose (A) | 7 of 11 (0) | 10 of 35 (6) | 8 of 17 (0) | | | | |

33 seconds, which occurred when analyzing `libcurl` and its three debloated variants.

The most time-consuming component of GSA is its gadget source identification feature, which requires recovering the binary's CFG with angr. This feature can be optionally disabled, significantly reducing the time required by GSA. When this feature is disabled, GSA is able to analyze `libcurl` and its three debloated variants in less than 40 seconds.

## 6. Results

In this section, we present the results generated by GSA across our debloating scenarios. Our results demonstrate both the utility of our proposed metrics and the shortcomings of gadget count reduction.

### 6.1. Functional Gadget Set Expressivity

Table 2 contains functional gadget set expressivity data collected by GSA. Debloating was generally successful in reducing expressivity when used aggressively on simple software packages (CHISEL benchmarks). However, this was not the case in the larger, more complex packages and less aggressive debloating scenarios (CARVE benchmarks).

In two scenarios, debloating reduced the gadget count but did not decrease gadget set expressivity (*italicized* text). In these cases, gadget count reduction indicates a positive security impact, but the actual impact on security is neutral. Of greater concern are the four scenarios in which debloating increased in gadget set expressivity (**bold** text) by introducing new gadgets that satisfied previously unsatisfied computational classes. In all four scenarios, using gadget count reduction as a security metric indicates positive results, but fails to identify this negative security impact.

### 6.2. Functional Gadget Set Quality

Table 3 contains the functional gadget set quality data collected by GSA. Results according to this metric are also concerning. In general, debloating succeeded in reducing the number of quality gadgets in with respect to each attack method (ROP / JOP / COP); we observed an increase in quality gadgets for at least one attack method in only 3 of the 21 debloating scenarios.

However, debloating generally did not succeed at significantly reducing the average quality score of the quality gadgets identified within each package binary. We observed only two scenarios in which debloating resulted in a positive result, which we define as a significant decrease in average quality with respect to at least one attack type, and no significant increases in quality (an increase is considered significant if its magnitude is 0.1 or larger). Conversely, we observed a negative result in *more than half* of scenarios (11 of 21), defined as a significant increase in average quality with respect to at least one attack type (**bold** text in Table 3), and no significant decreases in quality. In fact, we observed two scenarios in which debloating had decidedly negative results, indicated by a significant increase in average gadget quality with respect to all three attack types. Once again, gadget count reduction fails to capture these negative impacts.

### 6.3. Special Purpose Gadget Availability

Table 4 contains special purpose gadget availability data collected by GSA. The results measured according to this metric were similarly mixed. A conclusively positive result (complete elimination of all special purpose gadgets of some type) was observed in only 8 of 21 debloating scenarios. In eight other scenarios, debloating did not reduce the availability of special purpose gadgets (*italicized* text). In four scenarios, gadget introduction caused an increase in the count of special purpose gadgets (**bold** text). In one instance, debloating introduced a special purpose gadget of a type that was not available in the original package, a conclusively negative result. As was the case with functional gadget set expressivity and functional gadget set quality, measuring special purpose gadget availability captured mixed and negative debloating results that were not observable using gadget count reduction data.

### 6.4. Gadget Locality

The results measured according to this metric were conclusively positive for all benchmarks. In all but two scenarios, debloating (and subsequent re-compilation) caused a change in the component instructions and/or the offset within the binary for every gadget, resulting in 0% gadget locality. In the two scenarios in which we observed local gadgets, conservative debloating of libmodbus and Mongoose, the number of local gadgets after debloating was very small (three and five gadgets respectively). In both cases, the number of local gadgets accounted for 0.5% or less of the total remaining gadgets. Such a clear positive result is encouraging, and not readily apparent from gadget count reduction data.

**Table 3: Quality Gadget Counts and Average Quality Score (and Reduction) by Attack Method (ROP / JOP / COP).**

| | Debloated Variant | Quality ROP Gadgets | Average ROP Gadget Quality | Quality JOP Gadgets | Average JOP Gadget Quality | Quality COP Gadgets | Average COP Gadget Quality |
|---|---|---|---|---|---|---|---|
| CARVE | libmodbus (C) | 228 (61) | **0.81 (0.10)** | 45 (1) | **1.13 (0.13)** | 22 (0) | **0.36 (0.5)** |
| | libmodbus (M) | 231 (58) | 0.86 (0.05) | **62 (-16)** | 1.21 (0.05) | **23 (-1)** | **0.71 (0.15)** |
| | libmodbus (A) | 205 (84) | **0.76 (0.14)** | 61 (-15) | **0.99 (0.27)** | **30 (-8)** | **0.62 (0.25)** |
| | Bftpd (C) | 246 (16) | 0.96 (-0.02) | 31 (13) | 0.79 (-0.09) | 12 (12) | 0.88 (-0.21) |
| | Bftpd (M) | 227 (35) | 0.93 (0.01) | 31 (16) | **0.58 (0.12)** | 22 (2) | 0.59 (0.08) |
| | Bftpd (A) | 197 (65) | 0.91 (0.03) | 13 (34) | **0.50 (0.20)** | 8 (16) | **0.57 (0.10)** |
| | libcurl (C) | 1800 (25) | 0.86 (0.00) | 2978 (89) | 0.60 (-0.01) | 2420 (91) | 0.52 (-0.02) |
| | libcurl (M) | 1619 (206) | 0.85 (0.01) | 2492 (575) | 0.58 (0.02) | 2043 (468) | 0.48 (0.02) |
| | libcurl (A) | 1331 (494) | 0.84 (0.03) | 1668 (1399) | 0.57 (0.02) | 1420 (1091) | 0.48 (0.03) |
| | Mongoose (C) | 436 (8) | 1.07 (0.02) | 99 (11) | 0.79 (0.08) | 37 (12) | **0.51 (0.29)** |
| | Mongoose (M) | 422 (22) | 1.07 (0.02) | 95 (15) | 0.91 (-0.04) | 37 (12) | 0.96 (-0.15) |
| | Mongoose (A) | 400 (44) | 1.05 (0.03) | **119 (-9)** | 0.82 (0.05) | 44 (5) | 0.81 (0.00) |
| CHISEL | bzip2 | 96 (55) | 1.30 (-0.20) | 41 (23) | **0.29 (0.39)** | 36 (11) | **0.26 (0.43)** |
| | chown | 99 (124) | 1.35 (-0.01) | 12 (49) | **0.29 (0.38)** | 7 (25) | **0.36 (0.27)** |
| | date | 100 (120) | **1.01 (0.14)** | 21 (8) | 0.69 (-0.38) | **16 (-3)** | 0.81 (-0.54) |
| | grep | 277 (106) | **1.13 (0.17)** | 61 (87) | 0.75 (-0.01) | 48 (66) | 0.63 (0.08) |
| | gzip | 90 (75) | 1.13 (0.00) | 15 (15) | 0.5 (0.05) | 13 (6) | 0.54 (0.04) |
| | mkdir | 66 (46) | 1.10 (0.05) | 6 (6) | **0.08 (0.46)** | 3 (0) | 0.00 (0.00) |
| | rm | 72 (155) | **1.23 (0.11)** | 13 (45) | 0.58 (-0.09) | 8 (37) | 0.63 (-0.04) |
| | tar | 123 (298) | **1.15 (0.44)** | 39 (303) | 0.85 (-0.08) | 28 (253) | 0.64 (0.07) |
| | uniq | 66 (78) | 1.23 (0.04) | 9 (4) | 0.39 (0.00) | 2 (8) | **0.0 (0.45)** |

**Table 4: Special Purpose Gadget Counts (and Reduction) by Gadget Set (excludes types not observed in any variant).**

| | Package Variant | Syscall Gadget | JOP Dispatcher Gadgets | JOP Data Loader Gadgets | JOP Trampoline Gadgets | COP Dispatcher Gadgets | COP Intra Stack Pivot Gadgets |
|---|---|---|---|---|---|---|---|
| CARVE | *libmodbus (C)* | *0 (0)* | *0 (0)* | *3 (-2)* | *0 (0)* | *0 (0)* | *0 (0)* |
| | *libmodbus (M)* | *0 (0)* | *0 (0)* | *1 (0)* | *0 (0)* | *0 (0)* | *0 (0)* |
| | *libmodbus (A)* | *0 (0)* | *0 (0)* | *1 (0)* | *0 (0)* | *0 (0)* | *0 (0)* |
| | Bftpd (C) | 0 (0) | 0 (0) | 9 (1) | 0 (0) | 0 (0) | 0 (0) |
| | Bftpd (M) | 0 (0) | 0 (0) | 1 (9) | 0 (0) | 0 (0) | 0 (0) |
| | Bftpd (A) | 0 (0) | 0 (0) | 1 (9) | 0 (0) | 0 (0) | 0 (0) |
| | **libcurl (C)** | *5 (-1)* | 7 (1) | 50 (1) | 0 (1) | 304 (15) | 4 (0) |
| | libcurl (M) | 0 (4) | 4 (4) | 41 (10) | 0 (1) | 287 (32) | 3 (1) |
| | libcurl (A) | 4 (0) | 3 (5) | 14 (37) | 1 (0) | 170 (149) | 2 (2) |
| | **Mongoose (C)** | *0 (0)* | *1 (-1)* | *6 (0)* | *0 (0)* | *0 (0)* | *0 (0)* |
| | *Mongoose (M)* | *0 (0)* | *0 (0)* | *6 (0)* | *0 (0)* | *0 (0)* | *0 (0)* |
| | *Mongoose (A)* | *0 (0)* | *0 (0)* | *6 (0)* | *0 (0)* | *0 (0)* | *0 (0)* |
| CHISEL | *bzip2* | *0 (0)* | *0 (0)* | *1 (0)* | *0 (0)* | *0 (0)* | *0 (0)* |
| | chown | 0 (4) | 0 (0) | 1 (0) | 0 (0) | 0 (0) | 0 (0) |
| | date | 0 (4) | 0 (0) | 1 (0) | 0 (0) | 0 (0) | 0 (0) |
| | grep | 0 (4) | 0 (0) | 1 (3) | 0 (0) | 0 (0) | 0 (0) |
| | *gzip* | *0 (0)* | *0 (0)* | *1 (0)* | *0 (0)* | *0 (0)* | *0 (0)* |
| | mkdir | 0 (0) | 0 (0) | 1 (0) | 0 (0) | 0 (0) | 0 (0) |
| | rm | 0 (4) | 0 (0) | 1 (0) | 0 (0) | 0 (0) | 0 (0) |
| | **tar** | 0 (2) | 0 (0) | 2 (-1) | 0 (0) | 0 (0) | 0 (0) |
| | uniq | 0 (4) | 0 (0) | 1 (0) | 0 (0) | 0 (0) | 0 (0) |

These results indicate that source code-based debloating methods are very effective at producing variants that are resistant to exploit re-use, even when debloating is conservative in nature. Other code-removing tools that operate at lower levels program representation (e.g. TRIMMER and RAZOR) are likely to see similar benefits, however the specific technical methods may lend themselves to producing debloated variants with higher gadget locality. Methods that utilize binary rewriting as opposed to code removal to debloat (e.g. TOSS, Piece-wise compilation/loading, and BINTRIMMER [2, 5, 27]) intentionally avoid disturbing the binary representation of the program to avoid recalculation of pointers and offsets. We expect that these methods would exhibit high gadget locality values. We leave the task of comparing gadget locality data between different debloating techniques to future work.

## 7. Case Studies

In this section, we demonstrate through two case studies that our metrics can be used to achieve meaningful security-oriented goals. First, we use GSA to mitigate unexpected negative side effects that occurred when conservatively debloating `libcurl`. We use a process which we call *iterative debloating* to use information reported by GSA to help adjust our debloating specification to generate a new variant that debloats fewer features, but does not suffer the negative effects identified in the first debloating round. In the second case study, we use GSA to conduct a side by side comparison of CARVE and CHISEL to determine which was more effective at debloating `chown` v8.2. We use both the gadget count reduction metric, as well as our proposed metrics to show they provide a more meaningful comparison.

### 7.1. Case Study 1: Iterative Debloating

We observed a number of negative security impacts when conservatively debloating `libcurl`:

- Increased the gadget set expressivity with respect to ASLR-Proof practical ROP exploits
- Increased the gadget set expressivity with respect to simple Turing-completeness
- Increased the number of system call gadgets

In order to mitigate these effects, we attempted to find a less aggressive debloating configuration that does not suffer these negative side effects. In this scenario, GSA was not able to identify the containing function for the newly introduced gadgets. In the absence of this information, we used alternate heuristics to decide what adjustments to make to mitigate the negative effects. We observed the effects debloating different sets of features had on other variants of `libcurl`; These negative effects were not observed in both the aggressive and moderate debloating scenarios. By analyzing the feature sets selected for each variant, we determined that a second debloating attempt for `libcurl` (C) was likely to yield better results if we did not debloat support for two protocol families, SCP and RTSP.

We made this modification to the debloating specification, re-ran CARVE, built the new variant, and analyzed the new variant with GSA. The results were largely an improvement, as expected:

- Reduced gadget set expressivity with respect to ASLR-proof practical ROP exploits
- No change in gadget set expressivity with respect to simple Turing-completeness
- Decreased the number of system call gadgets

However, changing the debloating configuration did result in one negative impact. In the new variant, we observed an increase in the number of COP intra stack pivot gadgets by one gadget. While this is not an ideal result, we determined that it did not offset the improvements realized in this second round of debloating. We manually analyzed the newly introduced gadget and determined that it was functionally identical to another such gadget included in both the original package and our newly created variant; therefore, the newly introduced gadget does not significantly assist a hypothetical attacker.

### 7.2. Case Study 2: Comparing Debloating Tools

With a wide variety of debloating methods being presented in the literature, it is becoming more important to have useful metrics to meaningfully compare techniques to one another. To demonstrate that our proposed gadget-based metrics can be used to accomplish this, we used CARVE to debloat one of CHISEL's benchmark software packages, `chown` v8.2. We used the same specification for that was originally used by CHISEL's authors, which specifies the following features for debloating:

- Recursive Operations Options: traverse symbolic link from CLI (-H), traverse all symbolic links (-L), and do not traverse symbolic links (-P)
- Output Options: verbose reporting (-v), report changes only (-c), and suppress most output (-f)

We scanned the original binary and both debloated binaries with GSA and analyzed the results, which are displayed in Table 5. CHISEL achieved much higher gadget reduction (65.1% versus 25.7%) than CARVE, however our metrics tell a different and more complete story.

In each column of Table 5, the values of the debloating technique that performed better are in **bold** text. With respect to functional gadget set expressivity, CHISEL outperformed CARVE. CHISEL reduced the expressivity of the gadget with respect to two expressivity levels, while CARVE increased expressivity with respect to two levels. With respect to gadget quality, CHISEL was more effective than CARVE at eliminating quality ROP gadgets and reducing the average quality of ROP gadgets. However, CARVE outperformed CHISEL by a wide margin with respect to JOP and COP gadgets, reducing a larger number of quality gadgets and greatly reducing the average quality of the remaining gadgets. Chisel, on the other hand, did reduce the numbers of quality JOP and COP gadgets, but ended up increasing the average quality rather than reducing it.

Both techniques had identical results with respect to the availability of Special Purpose gadgets. CARVE and CHISEL both eliminated the four system call gadgets present in the original binary and left a single JOP Data Loader remaining. Also, both techniques achieved 0% gadget locality.

Table 5: Comparing the Effectiveness of Debloating `chown` v8.2 with CHISEL and CARVE.

| Package Variant | Total Gadgets | Functional Gadget Set Expressivity | | | Functional Gadget Set Quality | | | S.P. Gadget Categories Available | Gadget Locality |
|---|---|---|---|---|---|---|---|---|---|
| | | Practical ROP Exploit Classes | ASLR-Proof ROP Classes | Simple Turing Complete Classes | Quality ROP Gadgets Average ROP Quality | Quality JOP Gadgets Average JOP Quality | Quality COP Gadgets Average COP Quality | | |
| Original | 502 | 3 of 11 | 9 of 35 | 4 of 17 | 223<br>1.33 | 61<br>0.67 | 32<br>0.63 | 2 of 10 | N/A |
| CHISEL | **175 (65.1%)** | **3 of 11 (0)** | **5 of 35 (4)** | **2 of 17 (2)** | **99 (124)<br>1.34 (-0.01)** | 12 (49)<br>0.29 (0.38) | 7 (25)<br>0.357 (0.26) | 1 of 10 (1) | 0.0% |
| CARVE | 373 (25.7%) | 5 of 11 (-2) | 13 of 35 (-4) | 4 of 17 (0) | 132 (91)<br>0.92 (0.4) | **7 (54)<br>1.85 (-1.19)** | **6 (26)<br>2.12 (-1.54)** | 1 of 10 (1) | 0.0% |

Judging by gadget count reduction alone, CHISEL outperformed CARVE at this task by a wide margin. By contrast, our metrics reveal that CHISEL only clearly outperformed CARVE in one dimension. Further, CARVE was more successful at reducing the number of useful JOP and COP gadgets and their average quality, and performed equally well with respect to our other metrics. This disparity is further evidence of the inherent problems with gadget count reduction as a software debloating metric.

### 7.3. Discussion

Our case studies highlight several important consequences of the unpredictable nature of gadget introduction worthy of further discussion. First, it may not be possible to debloat packages with code-removing debloaters without incurring some negative impacts. Good security-oriented metrics should highlight these impacts and support the user in making informed tradeoffs when debloating, including choosing not to debloat.

Second, debloating does not have a linear relationship with security improvement. Techniques that debloat a larger portion of a package are not necessarily more effective at improving security than those that debloat less. Research into code-removing debloating techniques and tools should consider the problem of debloating in manner that minimizes or mitigates the negative side effects of gadget introduction.

## 8. Future Work

We have identified several areas worthy of future exploration. First, our proposed metrics are intended to generate discussion and research into other viable security-oriented metrics for software debloating, such as measures specific to particular exploit subtypes. Additionally, there is a need for new analysis tools that calculate gadget set expressivity data. Immediate needs raised by this work include gadget set expressivity analysis for JOP and COP gadgets, as well as expressivity analysis for arbitrary length gadgets. Finally, we intend to extend GSA with an optional feature to automatically analyze a variant's dynamically linked libraries.

## 9. Conclusion

We presented examples across a variety of debloating scenarios demonstrating the flaws inherent to the gadget count reduction metric. Despite achieving sizeable gadget count reductions, our scenarios revealed that debloating can introduce new gadgets, including gadgets of high value like special purpose gadgets. We proposed four new security-oriented metrics, functional gadget set expressivity, functional gadget set quality, special purpose gadget availability, and gadget locality to replace the gadget count reduction metric. We demonstrated that these metrics overcome the limitations of gadget count reduction by identifying when the side effects of debloating negatively impact security. Finally, we showed the practicality of these metrics by introducing our tool for calculating these metrics, GSA, and using it in two realistic case studies.

### Technical Artifact Availability

GSA, the tools it depends on (extended version of Gality and a publicly available microgadget scanner), and a selected sample of binaries analyzed in this work have been made publicly available to the community at:

https://github.com/michaelbrownuc/GadgetSetAnalyzer

https://github.com/michaelbrownuc/gality

### Acknowledgements

We would like to thank all of our reviewers for their helpful feedback. We also thank Andrei Homescu and his co-authors for allowing us to use their microgadget scanner in this work. Finally, we thank Joshua Kassab for his assistance in gathering our experimental data.

# A. Appendix – Mapped Features for CARVE Benchmarks

The following tables detail which features were mapped for each benchmark, and which were debloated in each scenario. For brevity, the complete list of fine-grained debloatable features in the package are condensed into the categories in the leftmost column. Features removed by CARVE are marked with an X in that scenario's column.

Table A.1: Debloated features per scenario for `libmodbus`.

| Debloatable Feature | Conservative Scenario | Moderate Scenario | Aggressive Scenario |
|---|---|---|---|
| RTU Read Operations | X | | X |
| RTU Write Operations | X | | X |
| RTU Raw Operations | X | | X |
| TCP (IPv4) Read Operations | | X | X |
| TCP (IPv4) Write Operations | | X | X |
| TCP (IPv4) Raw Operations | | X | X |
| TCP (IPv4/6) Read Operations | | X | |
| TCP (IPv4/6) Write Operations | | X | X |
| TCP (IPv4/6) Raw Operations | | X | X |

Table A.2: Debloated features per scenario for `Bftpd`.

| Debloatable Feature | Conservative Scenario | Moderate Scenario | Aggressive Scenario |
|---|---|---|---|
| Admin Commands | X | X | X |
| Read Commands | | | |
| Write Commands | | | X |
| Directory Commands | | X | X |
| Server Config Commands | | X | X |
| Miscellaneous Commands | | X | X |
| Server Info Commands | | | X |

Table A.3: Debloated features per scenario for `libcurl`.

| Debloatable Feature | Conservative Scenario | Moderate Scenario | Aggressive Scenario |
|---|---|---|---|
| Uncommon API Elements | X | X | X |
| HTTP | | | X |
| HTTPS | | | X |
| RTSP | X | X | X |
| FTP Read Commands | | | X |
| FTP Write Commands | | X | X |
| FTPS | | X | X |
| Telnet | X | X | X |
| LDAP | | X | X |
| TFTP | X | X | |
| IMAP | | X | X |
| SMB | | X | X |
| SMTP | | X | X |
| POP3 | | X | X |
| RTMP | X | X | X |
| File | X | X | X |
| Gopher | X | X | X |
| Dict | X | X | X |
| SCP | X | X | X |
| SFTP | | X | X |

Table A.4: Debloated features per scenario for `mongoose`.

| Debloatable Feature | Conservative Scenario | Moderate Scenario | Aggressive Scenario |
|---|---|---|---|
| Threads API | | | X |
| Broadcast API | X | | |
| IPV6 | X | | X |
| COAP | | | X |
| DNS | | X | X |
| HTTP Server | X | | |
| HTP CGI | X | | |
| Websocket | X | | |
| HTTP Client | | | X |
| Websocket | | | |
| MQTT | | | X |
| SNTP | X | | X |

# B. Appendix – Gality (Extended Version) Gadget Elimination Criteria and Side Effect Scoring

**Table B.1: Gality (Extended Version) Gadget Elimination Criteria.**

| Attack Category | Elimination Criteria Description |
|---|---|
| ROP | The gadget contains a non-static assignment to the stack pointer register. |
| ROP | The gadget contains a large (>16 bytes) or unaligned (%4) offset. |
| Syscall | The gadget is not ROP / JOP / specific (currently this means it is a system call gadget). |
| JOP / COP | The gadget ends with direct jump. Occurs due to a bug in ROPgadget. |
| JOP / COP | The indirect jump / call target register is derefenced and modified by an offset. |
| ROP / JOP / COP | The first instruction in the gadget is not within the class of useful gadgets. (pop, push, add, sub, adc, dec, inc, neg, not, mov, sbb, xchg, xor) |
| ROP / JOP / COP | The gadget contains a non-static assignment to the gadget's value destination register. |

**Table B.2: Side Effects Detected by Gality (Extended Version) and their Associated Scores.**

| Attack Category | Score Increase | Side Effect Description |
|---|---|---|
| ROP | 2.0 | A data move or arithmetic operation is performed on the stack pointer register. |
| ROP | 3.0 | A shift and rotate operation is performed on the stack pointer register. |
| ROP | 4.0 | The stack pointer register is the target of a move or exchange operation. |
| ROP | 2.0 | Gadget contains a leave instruction. |
| ROP | 2.0 | Cumulative stack pointer offsets in the gadget are negative. |
| ROP | 1.0 | The stack pointer register is the target of a pop operation. |
| JOP / COP | 2.0 | A data move or arithmetic operation is performed on the indirect jump/call target register. |
| JOP / COP | 3.0 | A shift and rotate operation is performed on the indirect jump/call target register. |
| JOP / COP | 2.0 | An exchange operation is performed on the indirect jump/call target register. |
| ROP / JOP / COP | 2.0 | Gadget contains a conditional branch instruction. |
| ROP / JOP / COP | 1.0 | A data move or arithmetic operation is performed on the gadget's value destination register. |
| ROP / JOP / COP | 1.5 | A shift and rotate operation is performed on the gadget's value destination register. |
| ROP / JOP / COP | 0.5 | A data move, arithmetic, or shift and rotate operations is performed on a potential value-carrying register. |